\documentclass[universe,communication,submit,moreauthors,pdftex,10pt,a4paper]{mdpi}
\usepackage{amsmath}
\firstpage{1} 
\articlenumber{x}
\doinum{10.3390/------}
\pubvolume{xx}
\pubyear{2018}
\copyrightyear{2018}
\history{Received: date; Accepted: date; Published: date}

\Title{Bayesian Analysis for Extracting Properties of the Nuclear Equation of State from Observational Data including Tidal Deformability from GW170817}

\Author{Alexander Ayriyan$^{1}$\orcidA{}, David Alvarez-Castillo$^{2}$\orcidB{}, David Blaschke$^{2,3,4}$\orcidC{}*, Hovik Grigorian$^{1,5}$\orcidD{}}

\AuthorNames{Alexander Ayriyan, David Alvarez-Castillo, David Blaschke, Hovik Grigorian}

\address{%
$^{1}$ \quad Laboratory for Information Technologies,
	Joint Institute for Nuclear Research,
	Joliot-Curie street 6,
	141980 Dubna, Russia\\
$^{2}$ \quad Bogoliubov Laboratory for Theoretical Physics,
	Joint Institute for Nuclear Research,
	Joliot-Curie street 6,
	141980 Dubna, Russia\\
$^{3}$ \quad Institute of Theoretical Physics, 
	University of Wroclaw, 
	Max Born place 9, 
	50-204 Wroclaw, Poland\\
$^{4}$ \quad National Research Nuclear University (MEPhI),
	Kashirskoe Shosse 31,
	115409 Moscow, Russia\\
$^{5}$ \quad Department of Physics, 
	Yerevan State University, 
 	Alek Manukyan Str. 1, 
 	0025 Yerevan, Armenia}

\corres{Correspondence: $^{*}$~david.blaschke@gmail.com}

\abstract{
We develop a Bayesian analysis method 
for selecting the most probable equation of state under a set of constraints from compact star physics, which now include the tidal deformability from GW170817.
We apply this method for the first time to a two-parameter family of hybrid equations of state that is based on realistic models for the hadronic 
phase (KVORcut02) and the quark matter phase (SFM$\alpha$) which produce a third family of hybrid stars in the mass-radius diagram.
One parameter ($\alpha$) characterizes the screening of the string tension in the string-flip model of quark matter while the other ($\Delta_P$) belongs to the mixed phase construction that mimics the thermodynamics of pasta phases and includes the Maxwell construction as a limiting case for $\Delta_P=0$.
We present the corresponding results for compact star properties like mass, radius and tidal deformabilities and use empirical data for them in the newly developed 
Bayesian analysis method to obtain the probabilities for the model parameters within their considered range. 
}


\usepackage{lipsum}
\usepackage{graphicx,pstricks,epsfig,rotating,amsmath,amssymb}

\newcommand{\bea}{\begin{eqnarray}}
\newcommand{\eea}{\end{eqnarray}}
\newcommand{\apgt} {\ {\raise-.5ex\hbox{$\buildrel>\over\sim$}}\ }

\preto{\abstractkeywords}{\nolinenumbers}

\begin{document}


\section{Introduction}

The determination of the equation of state (EoS) of compact star (CS) interiors is a subject of active research by means of both laboratory measurements and astronomical observations. 
One of the most urgent questions concerns the possible existence of deconfined quark matter in CS cores, where matter densities can exceed by several times the nuclear saturation value, $n_0=0.15$ fm$^{-3}$, the typical density in large atomic nuclei. 
Hybrid compact stars have an inner core composed of quark matter surrounded by an outer core of nuclear matter. 
Of particular interest is the CS mass twin (MT) phenomenon \cite{Glendenning:1998ag}: when a pair of stars has the same 
gravitational mass but different radii, with the larger star being composed only of hadronic matter and the smaller one being a hybrid star with a quark matter core. 
The presence of MTs is synonymous to the existence of a third family \cite{Gerlach:1968zz} of CS in the mass-radius ($M-R$) diagram.  
It requires the CS EoS to feature a strong first order phase transition from hadron to quark matter~\cite{Alford:2013aca,Alvarez-Castillo:2013cxa,Benic:2014jia,Blaschke:2015uva,Alvarez-Castillo:2017qki,Paschalidis:2017qmb}, where the latent heat of the transition fulfils the Seidov criterion \cite{1971SvA....15..347S}.

Should it turn out that the mass twin phenomenon can be discovered in CS observations this would give indirect evidence for the existence of a critical endpoint in the QCD phase diagram \cite{Alvarez-Castillo:2013cxa,Blaschke:2013ana,Benic:2014jia}
which is sought for in relativistic heavy-ion collisions, so far without being conclusive.
On the other hand, astronomical observations may provide measurements of masses and/or radii which are relevant for constraining the nuclear EoS via the one-to-one relationship provided by the Tolman-Oppenheimer-Volkoff (TOV) equations~\cite{Tolman:1939jz,Oppenheimer:1939ne}  which govern the general relativistic hydrodynamic stability of CS matter under its own gravity. 
Neutron star masses are accurately measured by monitoring binary pulsar dynamics whereas the radii were determined so far with big uncertainties.
The situation has changed with the first observation of the gravitational wave signal from the inspiral phase of the binary CS merger 
GW170817~\cite{TheLIGOScientific:2017qsa} which allowed to constrain the tidal deformability of both merging stars and thus their mass and 
radius~\cite{Annala:2017llu,Bauswein:2017vtn,Rezzolla:2017aly,De:2018uhw}. 

Consequently, theoretical approaches are required to quantitatively assess the most probable EoS out of a whole class obtained by varying intrinsic model parameters. 
One of the most powerful methods is the Bayesian analysis (BA) or Bayesian interpretation of probability that allows for estimation of model parameters based on prior knowledge, in this case the already collected measurements. Several works in this direction include BA based on observations of X-ray bursters~\cite{Steiner:2010fz} that unfortunately suffer from uncertainties in the stellar atmosphere composition modeling required in the interpretation of the observed signal or the more general approach of~\cite{Raithel:2017ity} that includes a generic multipolytrope EoS with priors that include experimental nuclear matter pressure values and that is able to support the most massive CS of about 2M$_{\odot}$ and reports an accuracy of about $30\%$ in pressure values. 
The recent approach of~\cite{Salmi:2018gsn} employs X-ray timing observations of accretion-powered millisecond pulsars, resulting in radius estimates of about $5\%$ if the CS mass is known. Moreover, a more stringent analysis~\cite{Margueron:2017lup} that incorporates the hypothesis of the Direct Urca cooling constraint~\cite{Blaschke:2016lyx} in addition to the afore mentioned measurements concludes that the neutron star radius has a value of $12.7\pm0.4$~km for masses ranging from 1 up to $2~M_{\odot}$.

Despite the fact that the recent detection of gravitational radiation from the inspiral phase of the binary CS merger GW170817 
has led to constraints on the tidal deformability of CS matter and to the discussion of the possibility that one or even both of the CS in the binary could be hybrid stars with quark matter interior and there would be a possibility to discover the mass twin phenomenon and thus a strong
first order phase transition in CS matter \cite{Alvarez-Castillo:2018pve,Most:2018hfd,Christian:2018jyd,Montana:2018bkb,Sieniawska:2018zzj}, these data have not yet been included in BA studies such as the ones already listed above. 
Thus, it is of great interest to update such BA studies by incorporating this new information to constrain the CS EoS.

In this work we consider the mass twin EoS and modifications to it in order to perform new Bayesian analysis calculations that, in addition to the previously used CS measurements in~\cite{Alvarez-Castillo:2016oln}, will include the GW170817 data as priors to assess the probability of model parameters. 
Our model choice is the KVOR EoS for hadronic matter together with the String-Flip approach for the deconfined quark regime.

\section{Hybrid EoS}

The hybrid EoS has been produced with the one-parameter replacement interpolation method (for a second order polynomial ansatz for $P(\mu)$)~\cite{Ayriyan:2017nby,Ayriyan:2017tvl,Abgaryan:2018gqp} which mimics the thermodynamic behavior of pasta phases in the coexistence region between the pure hadronic phase and the pure quark one.


The hadronic phase in this work is described by the well known KVOR equation of state~\cite{Kolomeitsev:2004ff} with a modification of stiffness as introduced in~\cite{Maslov:2015wba} and denoted as KVORcut2.
This particular version of the KVOR EoS model is the stiffest one presented in that work and allows to fulfill the condition  \cite{1971SvA....15..347S} for the latent heat of phase transition to quark matter, so that the disconnected hybrid star family at higher densities is possible.  

The quark phase is based on the String-Flip Model with the so called density functional~\cite{Kaltenborn:2017hus} including the available volume fraction $\Phi$,
\begin{equation}
\label{eq:exvol}
\Phi(n_\mathrm B) = {\mathrm e}^{- \alpha\ n_\mathrm{B}^2\ {\rm fm}^6},
\end{equation}
where the parameter $\alpha$ describes the effective density-dependence of the confining interaction, and it is varied here in the range $\left[0.1\dots 0.3\right]$. 
The value of this parameter is correlated with the critical density of the phase transition, as it is shown in the figures below. 
The larger $\alpha$, the lower the critical density and therefore the critical mass for the onset of the phase transition in the hybrid star. 
This systematics allows to calibrate the range of the model parameters with the observational data without contradicting the known properties of nuclear matter at saturation density.    

\begin{figure}[ht!]
\vspace{-8mm}
		\includegraphics[width=0.55\textwidth]{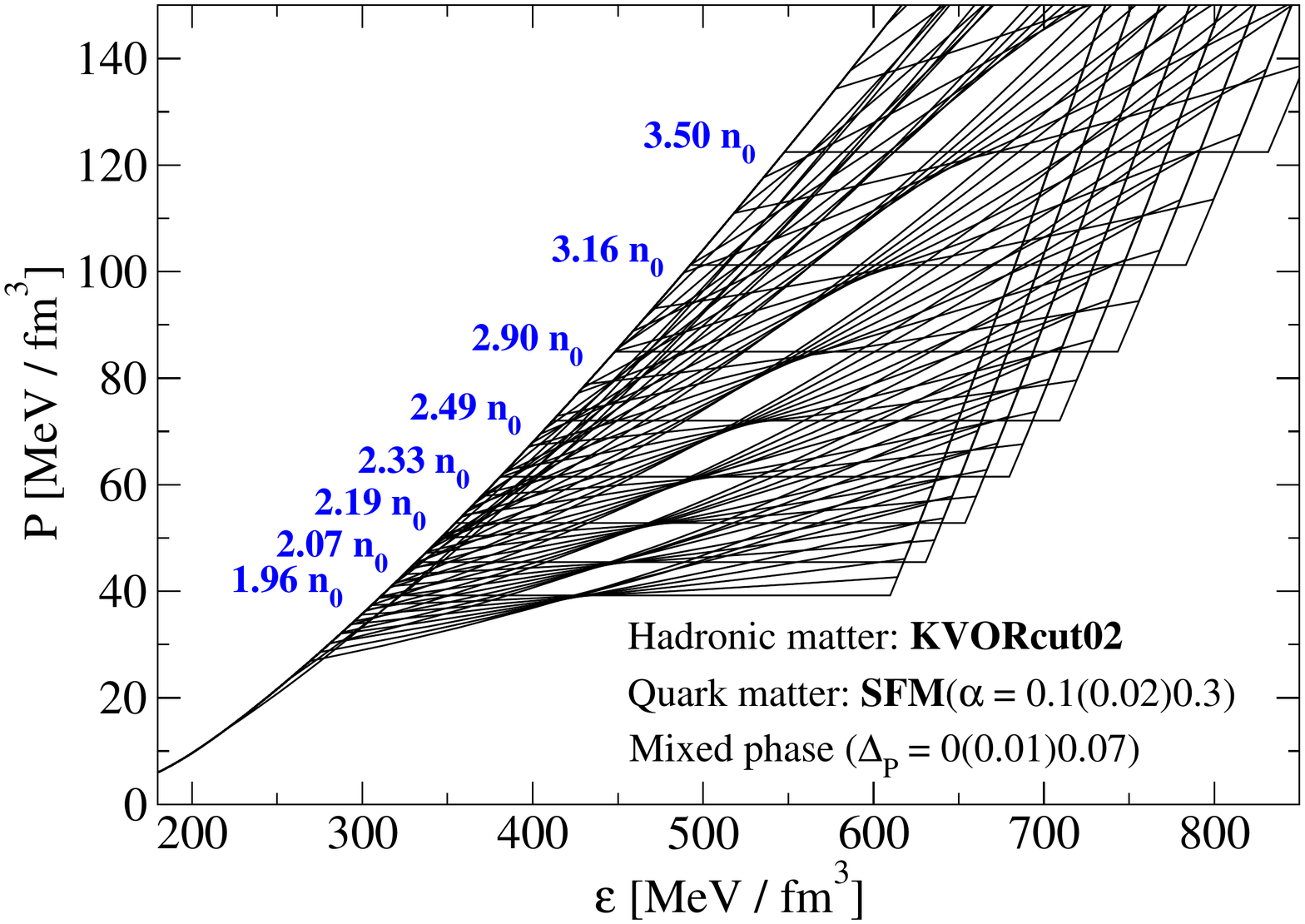} \hspace{-8mm}
		\includegraphics[width=0.55\textwidth]{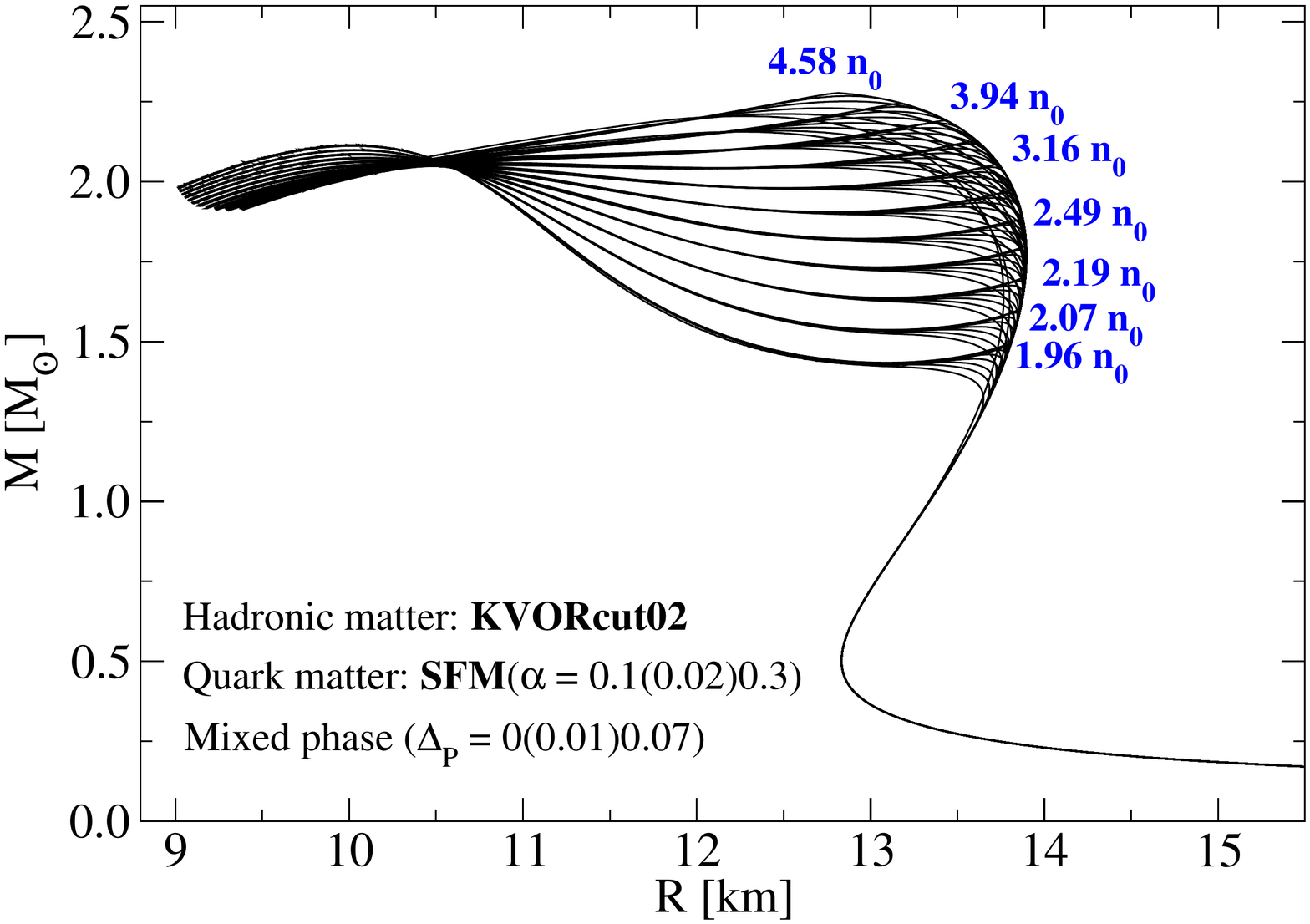}
	\caption{\label{fig:heos}
		The set of hybrid EoS obtained by the mixed phase construction for different values of $\Delta_P$ and $\alpha$ are shown in the left 		
		panel, while in the right panel the corresponding set of compact star sequences in the mass-radius diagram is shown. 
		The blue numbers show selected values of critical densities for the onset of the phase transitions obtained by a Maxwell construction.
		} 
\end{figure}

For the construction of the hybrid star EoS from the hadronic and quark matter EoS we employ here the mixed phase construction that is 
described in detail in Refs.~\cite{Ayriyan:2017tvl,Ayriyan:2017nby,Abgaryan:2018gqp}. 
It introduces the free parameter $\Delta_P=P(\mu_c)/P_c - 1$,
the pressure increment at the critical chemical potential $\mu_c$  relative to the critical pressure 
$P_c$ of the corresponding Maxwell construction, which is then obtained as limiting case for  $\Delta_P=0$.
The resulting family of hybrid EoS corresponds to the one described in \cite{Ayriyan:2017nby}, see Fig.~\ref{fig:heos}.
This way of modelling the phase transition mimics the possibility of so-called pasta phases characterized by different geometric structures 
in the coexistence region of the transition sufficiently well, as has been shown in \cite{Maslov:2018ghi}. 
In that reference it has also been demonstrated that the parameter $\Delta_P$ of the transition construction can be related to the value of the surface tension at the hadron 
to quark matter interface.
   
\section{Neutron star configurations}

\subsection{Mass and Radius}
The structure and global properties of compact stars are obtained by solving the TOV equations
\begin{eqnarray}
\label{eq:tov1}
\dfrac{dP(r)}{dr} &=& - \dfrac{G M( r)\varepsilon( r)}{r^2}\dfrac{\left(1+\dfrac{P( r)}{\varepsilon( r)}\right)\left(1+ \dfrac{4\pi r^3 P( r)}{M( r)}\right)}{\left(1-\dfrac{2GM( r)}{r}\right)}~,\\
\displaystyle \dfrac{dM( r)}{dr} &=& 4\pi r^2 \varepsilon( r)~,
\label{eq:tov2}
\end{eqnarray}
where $P(r)$, $\varepsilon(r)$ and $M(r)$ are the profiles of pressure, energy density and enclosed mass as a function of the distance $r$
from the center of the star. The radius $R$ of the star is obtained from the condition of vanishing pressure at the surface $P(R)=0$
and the gravitational mass of the star is $M=M(R)$.


\subsection{Tidal deformability}

\begin{figure*}[ht!]
		\includegraphics[width=0.55\textwidth]{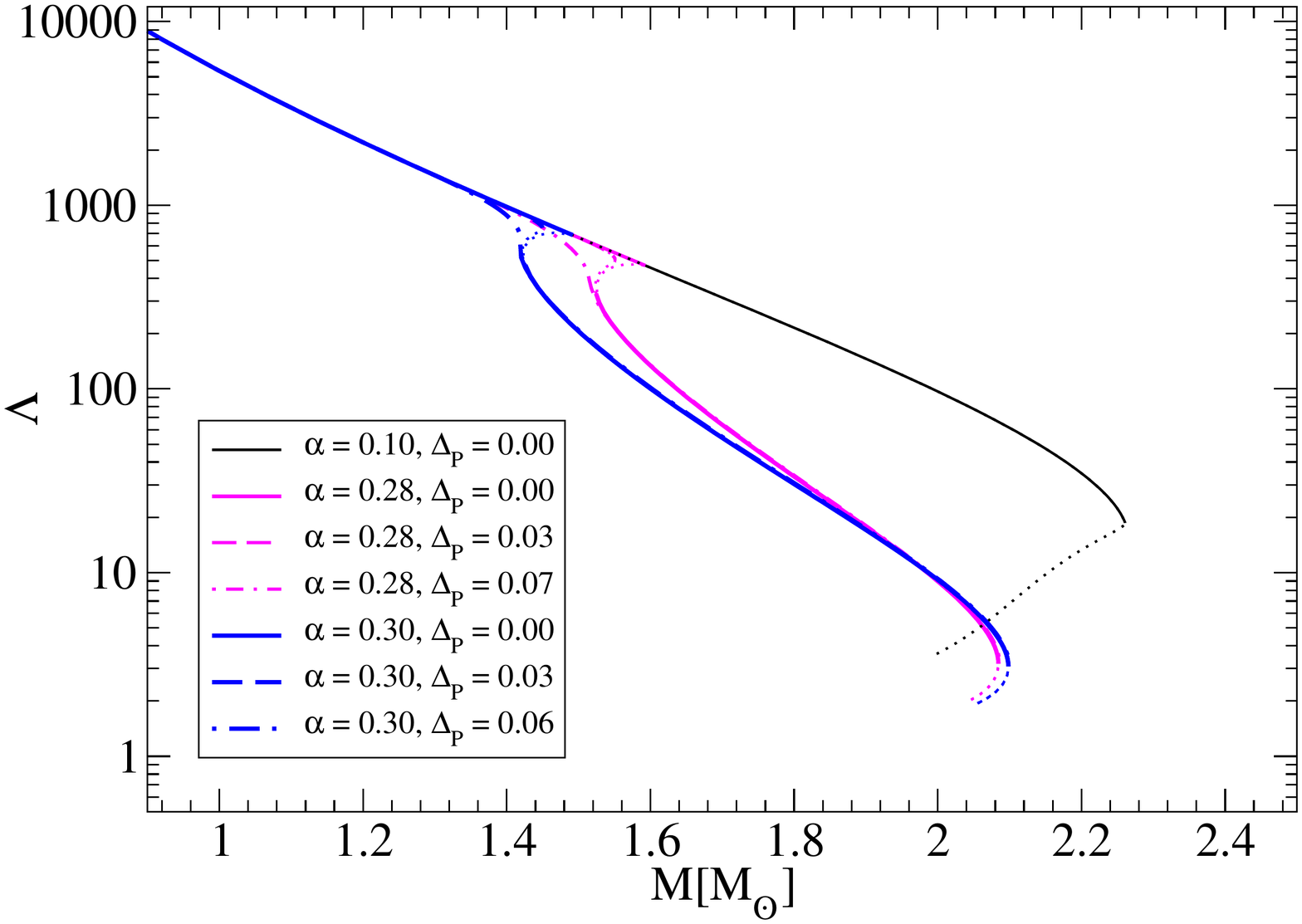}\hspace{-8mm}
		\includegraphics[width=0.55\textwidth]{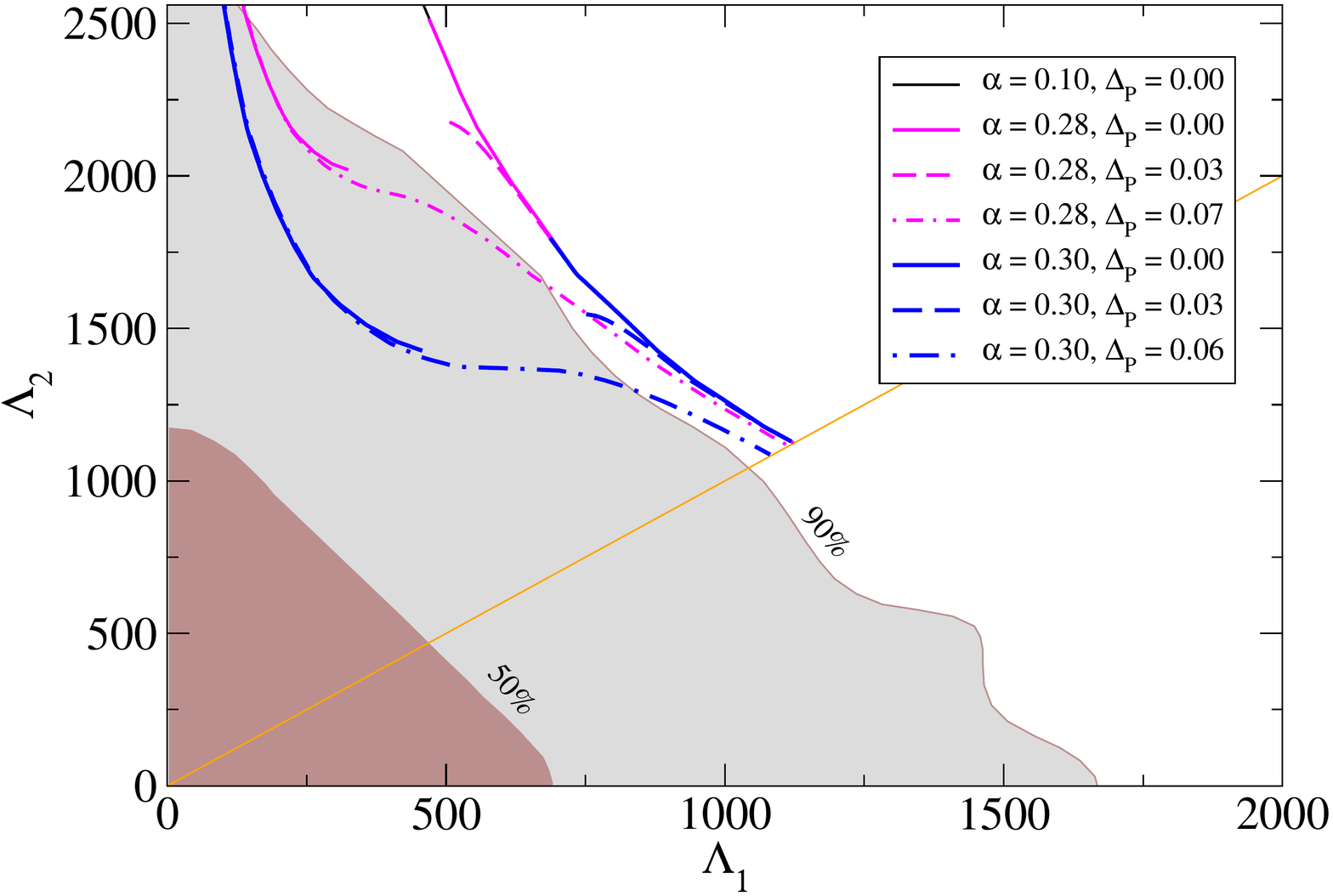}
	\vspace{-5mm}
	\caption{\label{fig:lambda}
		The dependence of the tidal deformability $\Lambda$ on the compact star mass (left panel) and the $\Lambda_1$--$\Lambda_2$ 			
		diagram for selected values of $\Delta_P$ and $\alpha$ (right panel).  
		Comparison with the 90\% confidence line of the LIGO-Virgo Collaboration for the low-spin prior of GW170817 
		\cite{TheLIGOScientific:2017qsa}  shows that if the hadronic EoS is as stiff as DD2$\underline{ }$p40 at least one of the 
		stars in the binary has to be a hybrid star in order to avoid a violation of the $\Lambda_1$--$\Lambda_2$ constraint.
		As can be seen from Fig.~17 of Ref.~\cite{Alvarez-Castillo:2018pve} the merger of two hybrid stars which is admissible when 
		the onset mass for the deconfinement transition is lowered, e.g., by increasing $\alpha$, would lead to the appearance of a new 
		branch in this diagram mimicking the pattern of a merger of two neutron stars with a rather soft nuclear EoS (like APR or SLy4). 
		} 
\end{figure*}

The recent observation of gravitational waves from the inspiral phase of the binary CS merger GW170817 gives an estimation of the tidal deformability of these stars, therefore we include also this information in the analyses. 
The tidal deformability of a star can be calculated from the Einstein equation for small elliptic deformation as described in~\cite{Hinderer:2009ca}. 
For the results of the calculation see Fig.~\ref{fig:lambda}.

\section{Bayesian inference for the EoS models}
\subsection{Vector of Parameters}
	The set of parameters of models could be represented in the parameter space with introduction of the vector of parameters, each vector is one fixed model from above described types of hadronic, quark phases and transition construction:   
\begin{equation}
	\label{pi_vec}
	\overrightarrow{\pi}_i = \left\{\alpha_{(j)},{\Delta_P}_{(k)}\right\},
\end{equation}
where $i = 0..N-1$ and $i = N_2\times j + k$~and~$j = 0..N_1-1$,~$k = 0..N_2-1$ and $N_1$ and $N_2$ are number of values of model parameters $\alpha$ and $\Delta_P$ correspondingly.

\subsection{Likelihood of a model for the $\Lambda_1-\Lambda_2$ constraint from GW170817 }

\begin{figure*}[ht!]
	\begin{center}$
		\begin{array}{cc}
		\includegraphics[width=0.49\textwidth]{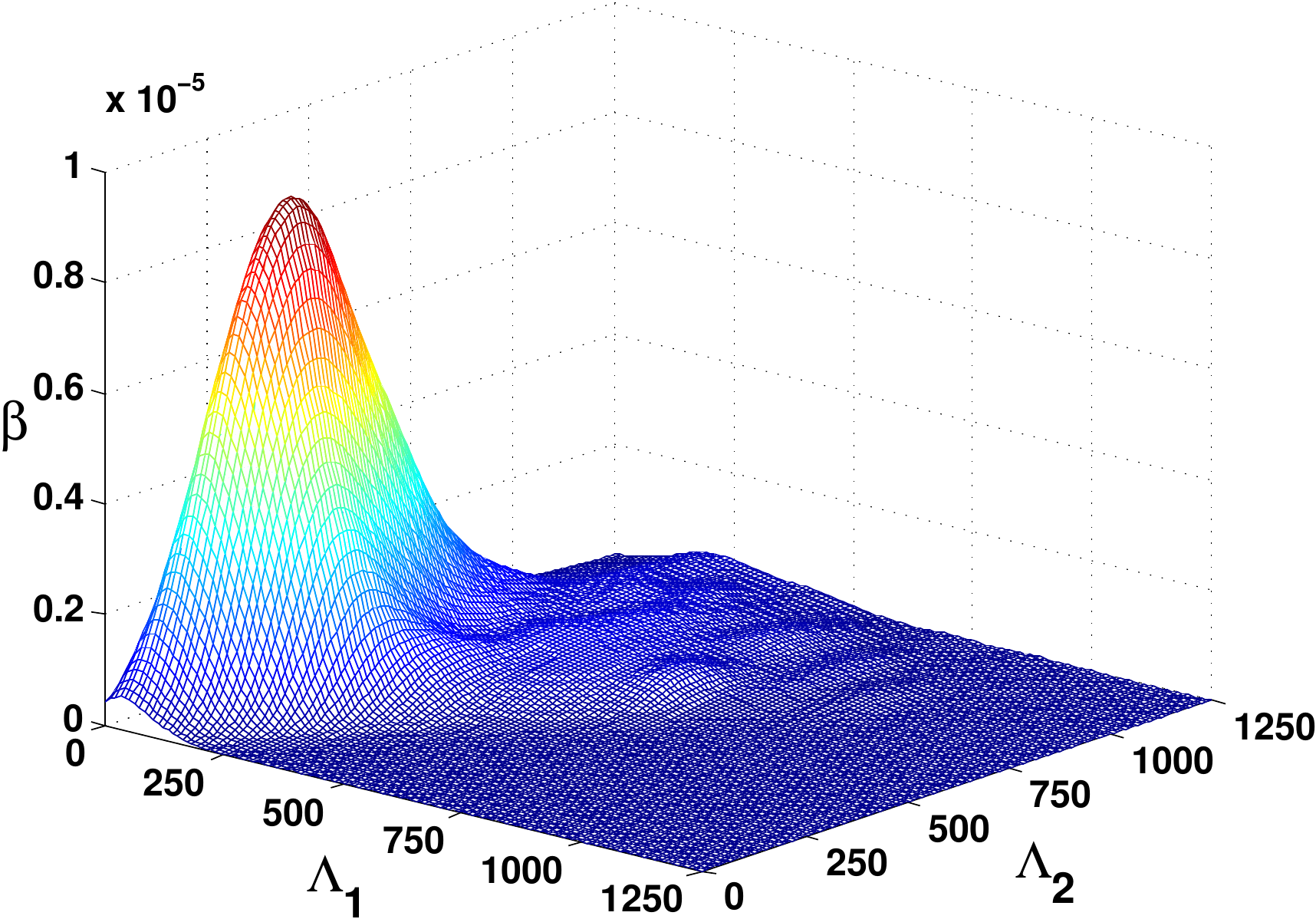} & 
		\includegraphics[width=0.46\textwidth]{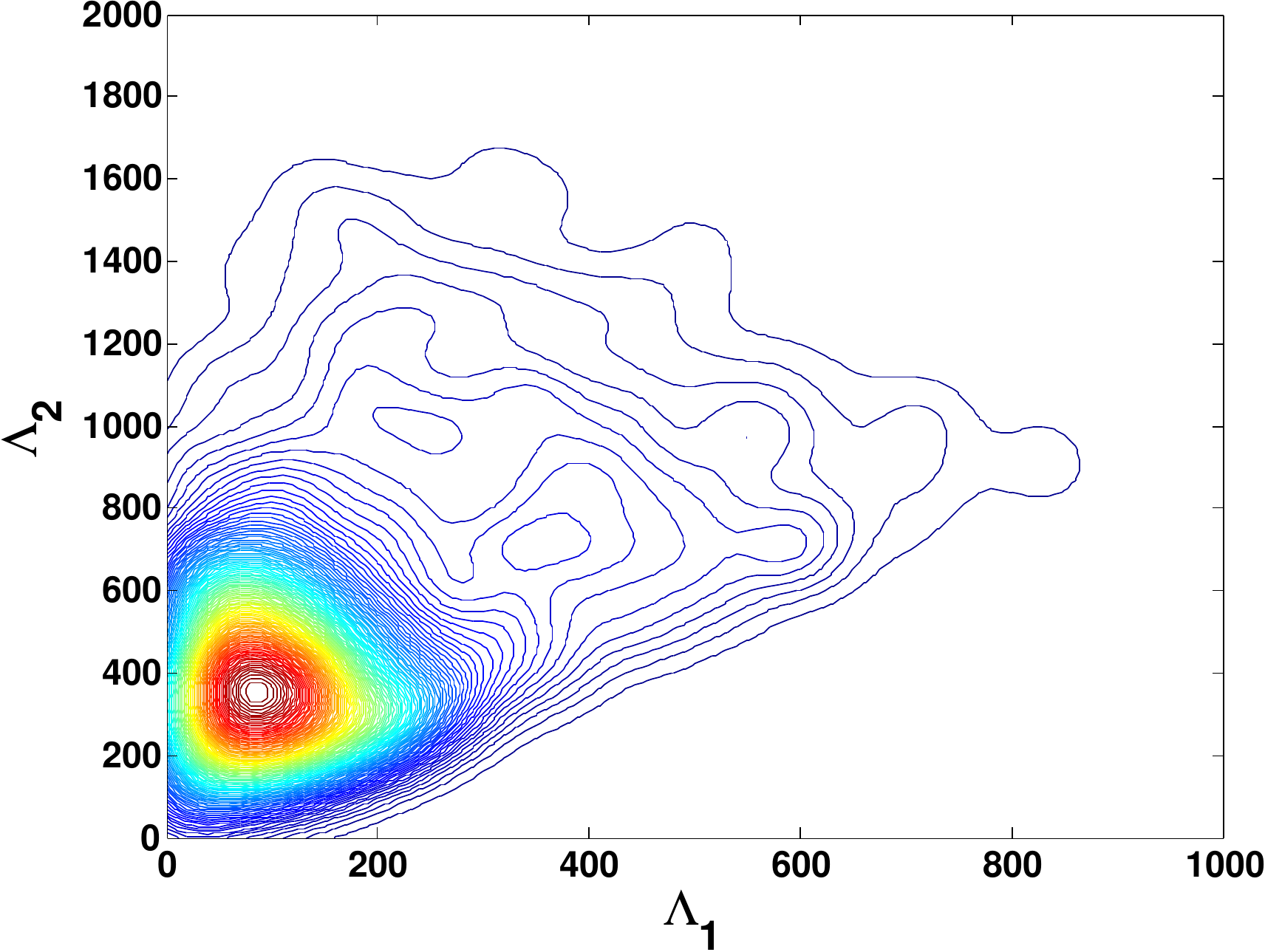}
		\end{array}$
	\end{center}
	\caption{The PDF reconstructed with $\Lambda_1$--$\Lambda_2$ data for GW170817 from the LIGO website \cite{LIGO} as a 3D graphics (left panel) and as a contour plot (right panel).} 
	\label{fig:L1L2_PDF} 
\end{figure*}

In order to implement the tidal deformability constraint on the compact stars EoS, reflected on the $\Lambda_1$--$\Lambda_2$ diagram that includes probability regions from GW170817 event~\cite{TheLIGOScientific:2017qsa,Abbott:2018exr}, we employ the definition of the probability as an integral over the probability distribution function (PDF)  
$\beta(\Lambda_1, \Lambda_2)$ by 
\begin{equation}
\label{eq:lhoodLL}
P\left(E_{GW}\left|\pi_i\right.\right) = \int_{l_{22}} \beta(\Lambda_1(\tau), \Lambda_2(\tau))d\tau, 
\end{equation}
when both stars in the binary merger belong to the second family branch of neutron stars (and $l_{22}$ is the corresponding path in the 
$\Lambda_1-\Lambda_2$ plane), or by
\begin{equation}
\label{eq:lhoodLLtwins}
P\left(E_{GW}\left|\pi_i\right.\right) = \int_{l_{22}} \beta(\Lambda_1(\tau), \Lambda_2(\tau))d\tau + \int_{l_{23}} \beta(\Lambda_1(\tau), \Lambda_2(\tau))d\tau, 
\end{equation}
in case the parameter vector $\pi_i$ corresponds to hybrid star equation of state with a third family of compact stars in the mass range relevant for the merger. 
Then $l_{23}$ is the path in the $\Lambda_1-\Lambda_2$ plane corresponding to a binary merger composed of a neutron star and a hybrid star. 
The parameter $\tau$ defines the position of a point on the paths $l_{22}$ and $l_{23}$ in equations \eqref{eq:lhoodLL}--\eqref{eq:lhoodLLtwins}.
Note that the PDF has been reconstructed by the method Gaussian kernel density estimation with $\Lambda_1-\Lambda_2$ data given at LIGO web-page \cite{LIGO}, see~fig.~\ref{fig:L1L2_PDF}.

\subsection{Likelihood of a model for the mass constraint}

The likelihood of the model is the conditional probability of the expected value of the possible maximum mass for the given model parameter vector:  
\begin{equation}
\label{eq:lhoodMass}
P\left(E_{A}\left|\pi_i\right.\right) = \Phi(M_i, \mu_A, \sigma_A),
\end{equation}
here $M_i$ is maximum mass of the given by $\pi_i$, and $\mu_A = 2.01~\mathrm{M_{\odot}}$ and $\sigma_A = 0.04~\mathrm{M_{\odot}}$ is the mass measurement~\cite{Antoniadis:2013pzd}.

\subsection{Posterior distribution}
The full likelihood for the given $\pi_i$ can be calculated as a product of all likelihoods, since the considered constraints are independent of 
each other
\begin{equation}
\label{eq:p_event}
	P\left(E\left|\overrightarrow{\pi}_{i}\right.\right)= \prod_{m} P\left(E_{m}\left|\overrightarrow{\pi}_{i}\right.\right).
\end{equation}
The posterior distribution of models on parameter diagram is given by Bayes' theorem
\begin{equation}
\label{eq:bayes}
	P\left(\overrightarrow{\pi}_{i}\left|E\right.\right)=\frac{P\left(E\left|\overrightarrow{\pi}_{i}\right.\right)P\left(\overrightarrow{\pi}_{i}\right)}{\sum\limits _{j=0}^{N-1}P\left(E\left|\overrightarrow{\pi}_{j}\right.\right)P\left(\overrightarrow{\pi}_{j}\right)},
\end{equation}
where $P\left(\overrightarrow{\pi}_{j}\right)$ is a prior distribution of a models taken to be uniform: $P\left(\overrightarrow{\pi}_{j}\right)=1/N$. 

\begin{figure*}[!ht]
		\includegraphics[height=0.43\textwidth]{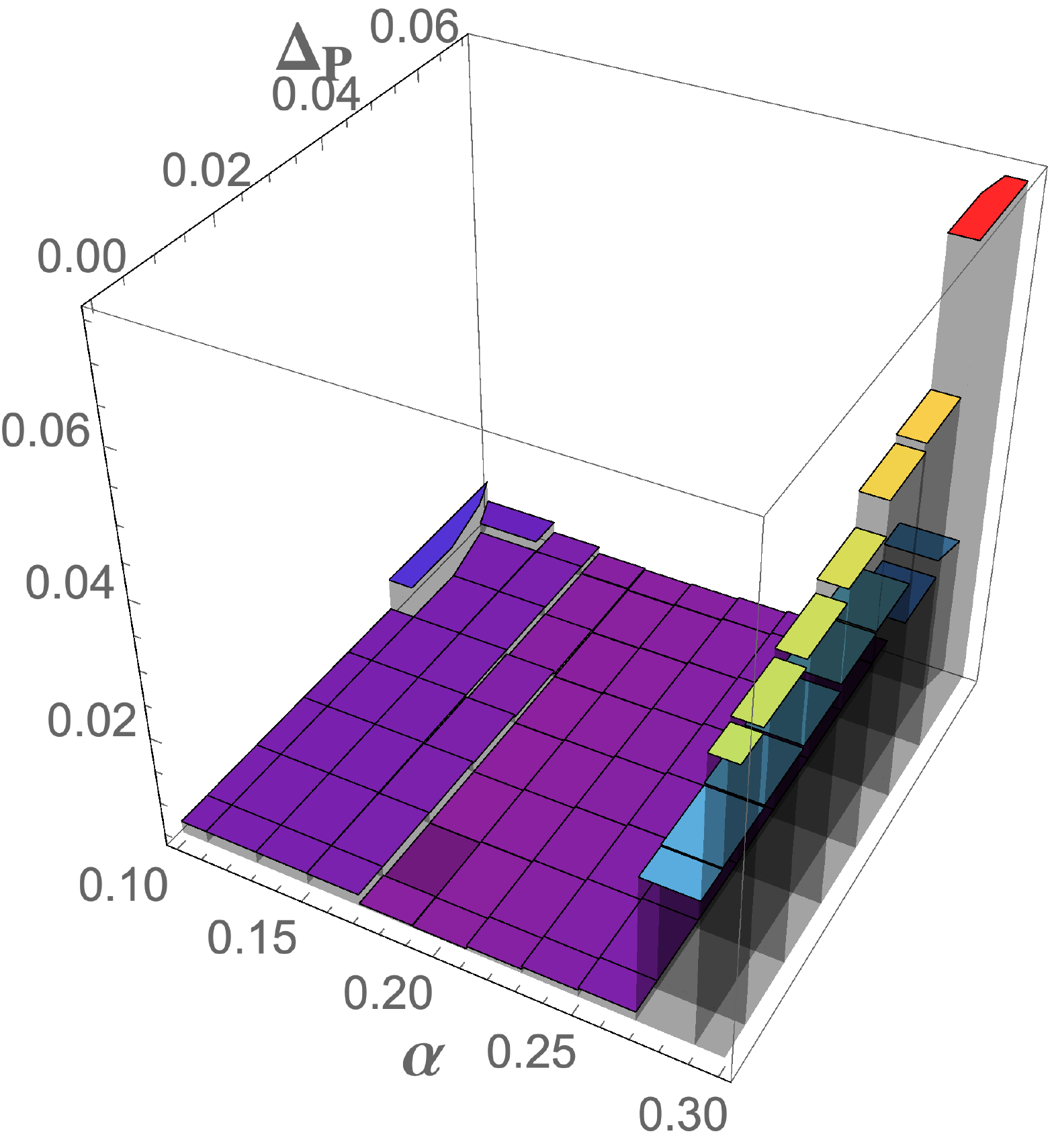} \hspace{-1mm}
	\includegraphics[height=0.5\textwidth]{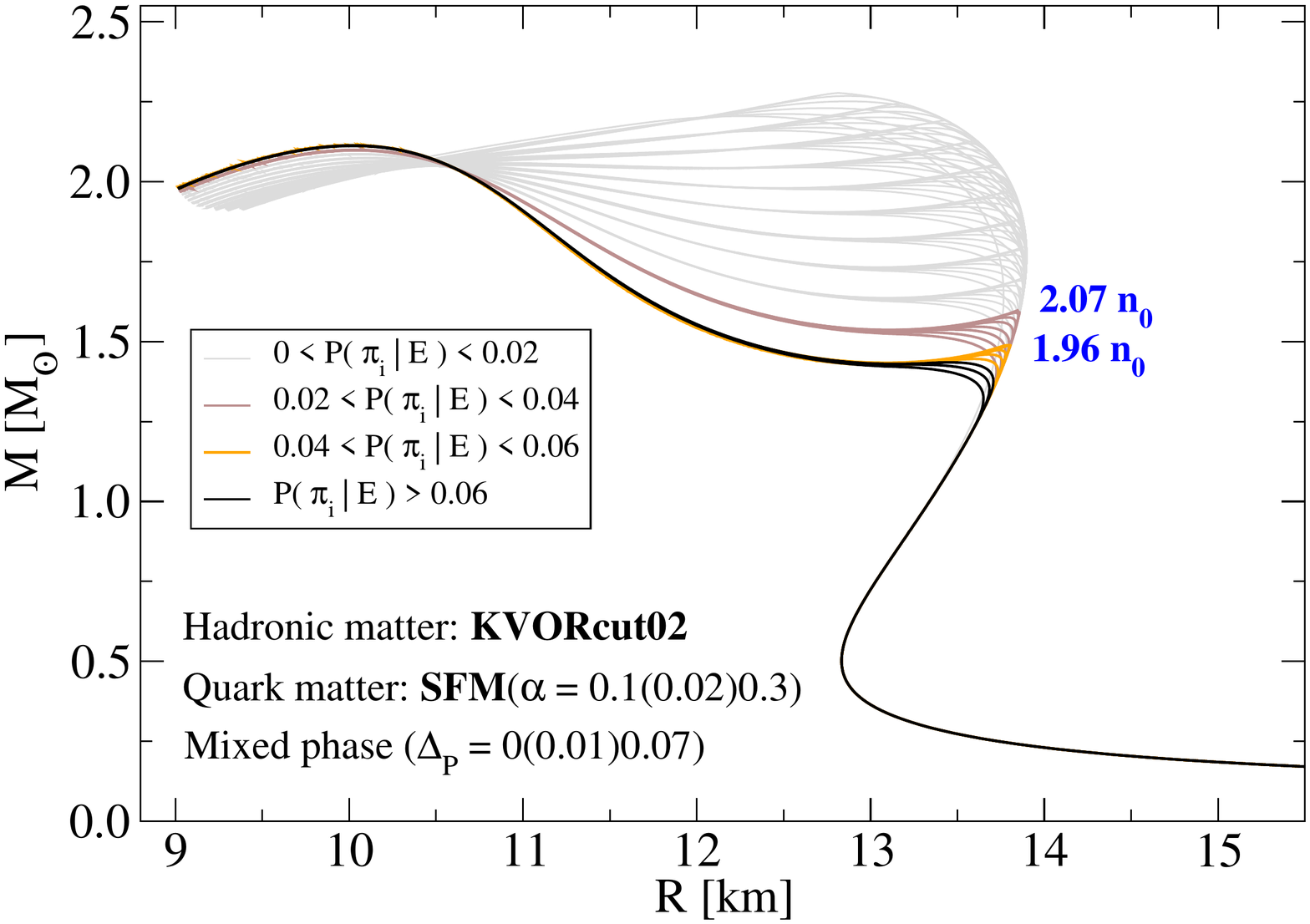}
	\caption{Left panel: The posterior distribution of models on the parameter space spanned by $\Delta_P$ and $\alpha$.
	Right panel: The compact star sequences in the mass-radius diagram labeled into four probability classes according to the results of the BA
	for the posterior distribution of the left panel. The grey, brown, orange and black lines show sequences for which the value of the posterior 
	probability exceeds the thresholds of 0.0, 0.02, 0.04 and 0.06, respectively.
	Due to the restriction to the model and parameter range used in Ref.~\cite{Ayriyan:2017nby}, the sequences with a lower onset mass for the 
	deconfinement transition are not accessed for which an interpretation of GW170817 as a merger of two hybrid stars would be possible. }  
	\label{fig:results} 
\end{figure*}
%

\section{Results}
The results of the Bayesian analysis are given in Fig.~\ref{fig:results}.
The most likely EoS are those with a strong mixed phase effect described by $\Delta_P  > 0.04$ and with a large screening parameter $\alpha > 0.28$,
at the limit of the parameter range considered here. For these parameter sets the phase transition and therefore the compactification of the 
hybrid star configuration occurs within the mass range that is relevant for GW170817. 
These results indicate that it should be worthwile to repeat this exploratory calculation with a wider set of parameters, in particular a larger 
screening parameter $\alpha$. 

\section{Conclusions}
We have developed a Bayesian analysis method for selecting the most probable equation of state under a set of constraints from compact star physics, 
which now include the tidal deformability from GW170817.
We have applied this method to a case study that employed the class of hybrid EoS introduced in  \cite{Ayriyan:2017nby} that allow for the existence of a third family 
of compact stars. 
We have investigated the requirements on future measurements to find the equivalent phenomenon of mass twins, i.e. two objects 
with the same mass but different, distinguishable radii. 
A mass - radius relation with two branches is mapped also on the $\Lambda_1-\Lambda_2$ diagram as it is shown here in Fig.~\ref{fig:L1L2_PDF} 
and other recent publications  \cite{Alvarez-Castillo:2018pve,Most:2018hfd,Christian:2018jyd,Montana:2018bkb,Sieniawska:2018zzj}. 
Since binary compact star mergers are expected to be observed by the LIGO-Virgo Collaboration at a rate of 1-10 events per year, we expect that with the observation 
of next merger events a binodal structure of the PDF in the $\Lambda_1-\Lambda_2$ plane could become apparent as a manifestation of the low-mass twin case with 
an onset of the third family branch below $\sim 1.3~M_\odot$ if such a branch exists in nature. 
A similar suggestion has been already proposed by Christian et al. \cite{Christian:2018jyd}.
Such an observation would support the existence of an early phase transition, around $2n_0$ for strong in-medium screening of the string tension.

\section{Acknowledgements}
We acknowledge discussions with K. Maslov on the hybrid star EoS.
A.A., D.B., and H.G. acknowledge support from the Russian Science Foundation under grant No. 17-12-01427 for the work described in 
sections 2 - 6. D. A.-C. is grateful for partial support from the Ter-Antonian - Smorodinsky program for collaboration between JINR and Armenian 
scientific institutions and from the Bogoliubov-Infeld program for collaboration between JINR and Polish Institutions.


\reftitle{References}




\end{document}